\documentclass{elsart}

\usepackage[latin1]{inputenc}
\usepackage{amssymb}

\newtheorem{The}{Theorem}[section]

\newtheorem{Pro}[The]{Proposition}
\newtheorem{Deff}[The]{Definition}
\newtheorem{Lem}[The]{Lemma}
\newtheorem{Rem}[The]{Remark}
\newtheorem{Exa}[The]{Example}

\newcommand{\fa}{\forall}

\newcommand{\Si}{\Sigma}
\newcommand{\Sio}{\Sigma^\omega}

\newcommand{\ra}{\rightarrow}
\newcommand{\hs}{\hspace{12mm}

\noi}

\newcommand{\Exab}{\begin{Exa}}
\newcommand{\Exae}{\end{Exa}}

\newcommand{\lra}{\leftrightarrow}

\newcommand{\ite}{\item}
\newcommand{\vp}{\varphi}

\newcommand{\orl}{ regular $\omega$-language}
\newcommand{\om}{\omega}
\newcommand{\nl}{\newline}
\newcommand{\noi}{\noindent}
\newcommand{\proo}{\noi {\bf Proof.} }
\newcommand {\ep}{\hfill $\square$}

\begin{document}
\begin{frontmatter}

\title{\bf On Decidability Properties of Local Sentences}
\author{Olivier Finkel}
\corauth[cor]{Corresponding author}
\ead{finkel@logique.jussieu.fr }

\address{ Equipe de Logique Math\'ematique \\CNRS  et 
Universit\'e Paris 7,  U.F.R. de Math\'ematiques \\  2 Place Jussieu 75251 Paris
 cedex 05, France.}

\begin{abstract} 
Local (first order) sentences, introduced
by Ressayre, enjoy very nice decidability properties, following from
some stretching theorems stating
some remarkable links between the finite and the infinite model theory of
these sentences
\cite{ress}.
Another stretching theorem of Finkel and Ressayre implies that one can
decide,
for a given local sentence $\vp$ and an ordinal $\alpha <\om^\om$, whether
$\vp$ has
a model of order type $\alpha$.
This result is very similar to B\"uchi's one who proved that the
monadic second order theory of the structure $(\alpha, <)$, for a
countable ordinal $\alpha$, is decidable. It is in fact
an extension of that result, as shown in \cite{loc} by considering the
expressive power
of monadic sentences and of local sentences over languages of words of
length $\alpha$.
The aim of this paper is twofold. We wish first to attract the reader's
attention
on these powerful decidability results proved using methods of model
theory and which
should find some applications in computer science and we prove also
here several additional results on local sentences.
\nl The first one is a new decidability result
in the case of local sentences whose function symbols are at most unary:
one can decide, for every {\it regular cardinal} $\om_\alpha$ (the
$\alpha$-th infinite 
cardinal), whether a local sentence $\vp$ has a model of order type
$\om_\alpha$.
\nl Secondly we show that this result can not be extended to the general
case. Assuming the
consistency of an inaccessible cardinal we prove that the set
of local sentences having a model of order type $\om_2$ is not determined by
the
axiomatic system $ZFC + GCH$, where $GCH$ is the generalized continuum
hypothesis.
\nl Next we prove that for all integers $n, p \geq 1$, 
if $n < p$ then the local theory of $\om_n$, i.e. the set
of local sentences having a model of order type $\om_n$,  
is recursive in the local theory of $\om_p$ and also in the local theory of $\alpha$ where 
$\alpha$ is any ordinal of cofinality $\om_n$.

\end{abstract}


\end{frontmatter}

\section{Introduction}

\noi A local sentence is a first order sentence
which is equivalent to a universal sentence and satisfies some semantic
restrictions:
closure in its models takes a finite number of steps.
Ressayre introduced local sentences in \cite{ress} and
established
some remarkable links between the finite and the infinite model theory of
these sentences
given by some stretching theorems.
Assuming that a binary relation symbol belongs to the signature of a local
sentence $\vp$
and is interpreted by a linear order in every model of $\vp$, the
stretching theorems state that the existence of some well ordered models of
$\vp$ is equivalent to
the existence of some finite model of $\vp$, generated by
some particular kind of indiscernibles, like special, remarkable or
monotonic ones.
Another stretching theorem of Finkel and Ressayre
establishes the equivalence between the existence of a model of order type
$\alpha$
(where $\alpha$ is an infinite ordinal $<\om^\om$) and the existence of a
finite model
(of another local sentence
$\vp_\alpha$) generated by $N_{\vp_\alpha}$ semi-monotonic indiscernibles
(where $N_{\vp_\alpha}$ is a positive integer depending on $\vp_\alpha$)
\cite{fr}.
\nl This theorem provides some decision algorithms which show the
decidability of the
following problem:
~~ {\bf $(P)$} ``For a given local sentence $\vp$ and an ordinal $\alpha
<\om^\om$, has $\vp$
a model of order type $\alpha$ ?"
\nl This last result is very similar to B\"uchi's one who proved that the
monadic second order theory of the structure $(\alpha, <)$, for a
countable ordinal $\alpha$, is decidable, \cite{bu62,tho,bs}.
B\"uchi obtained some decision algorithms by proving firstly that, for
$\alpha$-languages (languages of  infinite
words of length $\alpha$) over a finite alphabet,
definability by monadic second order sentences is equivalent to acceptance
by finite automata
where a transition relation is added for limit steps.
\nl We can compare the expressive power of monadic sentences and
of local sentences, considering languages defined by these sentences.
For each ordinal $\alpha < \om^\om$, an
$\alpha$-language over a finite alphabet $\Sigma$ 
is called local in \cite{ress,fr} (or also locally
finite
in \cite{loc,cloloc,toploc}) iff it is
defined by a second order sentence in the form
$\exists R_1\ldots\exists R_k \varphi$,
where $\varphi$ is local
in the signature $S(\varphi) =\{<, R_1,\ldots R_k, (P_a)_{a\in \Sigma}\}$, 
$R_1,\ldots R_k$ are relation or
function symbols, and, for each $a\in\Sigma$, $P_a$ is a unary predicate symbol.
\nl The class $LOC_\alpha$ of local $\alpha$-languages,
for $\om \leq \alpha < \om^\om$, is a strict extension of the class
$REG_\alpha$
of regular $\alpha$-languages, defined by monadic second
order sentences \cite{loc}.
Moreover this extension is very large. This can be seen by
considering the topological complexity of $\alpha$-languages and firstly of
$\om$-languages.
It is well known that all \orl s are boolean combinations of
${\bf \Si^0_{2}}$ Borel sets hence ${\bf \Delta^0_{3}}$ Borel sets,
\cite{tho,pp}.
On the other hand the class
$LOC_\om$ meets all finite levels of the Borel hierarchy, contains some
Borel sets of infinite rank and even some analytic but non Borel
sets, \cite{toploc}.
\nl Thus the decision algorithm for local sentences
provides in fact a very large extension,
for $\alpha <\omega^\omega$, of
Büchi's result about the decidability of the monadic second order theory of
$(\alpha, <)$.
Moreover, at least for $\alpha=\om$, the algorithm for local sentences is
of much lower complexity than the corresponding algorithm for monadic
second order sentences \cite{toploc}.
\nl We think that these powerful decidability results proved using methods
of model theory
should find some applications in computer science and that the study of
local sentences
could become an interdisciplinary subject for both model theory and computer
science
communities.
\nl So the aim of this paper is twofold: firstly to attract the reader's
attention on these
good properties of local sentences and their possible further applications;
secondly to
prove several new results on local sentences, described below.
\nl B\"uchi showed that for every ordinal $\alpha < \om_2$, where $\om_2$ is
the second
uncountable cardinal, the monadic theory of $(\alpha, <)$ is decidable. This
result cannot
be extended to $\om_2$. Assuming the existence of a weakly compact cardinal
(a kind of large
cardinal) Gurevich, Magidor and Shelah proved that the monadic theory of
$(\om_2, <)$ is not
determined by the set theory axiomatic system $ZFC$. They proved even much
more: for any
given $S \subseteq \om$ there is a model of ZFC where the monadic theory of
$(\om_2, <)$
has the Turing degree of $S$; in particular it can be non-recursive
\cite{gms}.
\nl Ressayre asked similarly for which ordinals $\alpha$ it is decidable
whether a given
local sentence $\vp$ has a model of order type $\alpha$. The question is
solved in \cite{fr}
for $\alpha < \om^\om$ but for larger ordinals the problem was still open.
\nl We firstly consider local
sentences whose function symbols are at most unary. We show that these
sentences satisfy
an extension of the stretching theorem implying new decidability properties.
In particular,
for each {\it regular} cardinal $\om_\alpha$ (hence in particular for each
$\om_n$ where $n$ is a positive integer), it is decidable
whether a local sentence
$\vp$ has a model of order type $\om_\alpha$. To know that this restricted
class $ LOCAL(1)$
of local sentences has more decidability properties is of interest
because it has already a great expressive power.
\nl Sentences in $ LOCAL(1)$ can define all regular finitary languages
\cite{ress}, and all the
quasirational languages forming a large class of context free languages
containing all
linear languages \cite{loc}.
\nl If we consider their expressive power over infinite words, sentences in
$ LOCAL(1)$ can define all \orl s \cite{loc}, but also some ${\bf
\Si^0_{n}}$-complete and
some ${\bf \Pi^0_{n}}$-complete Borel sets for every integer $n\geq
1$,
\cite{toploc}.
\nl Next we show that this decidability result can not be extended to local
sentences
having $n$-ary function symbols for $n\geq 2$.
Assuming the consistency of an inaccessible cardinal, we prove that
the local theory of $\om_2$ (the set of local sentences having a model of
order type $\om_2$)
is not determined by the system $ZFC + GCH$, where $GCH$ is the generalized
continuum hypothesis. This is also extended to
many larger ordinals.
\nl This result is obtained by showing that there is a local sentence which
has a
model of order type $\om_2$ if and only if there is a Kurepa tree, i.e. a
tree
of height $\om_1$ whose levels are countable and which has more than
$\om_1$ branches of length $\om_1$.
Kurepa trees have been much studied in set theory and their existence has
been shown to be
independent of $ZFC + GCH$, via the consistency of $ZFC + \mbox{`` there is
an inaccessible
cardinal"}$.
\nl It is remarkable that our proof
needs only the consistency
of an inaccessible cardinal which
is still a large cardinal but a very smaller cardinal than a weakly compact
cardinal.
This gives another indication of the great expressive power of local
sentences with regard to that of monadic sentences.
\nl We could still expect,
as Shelah did in \cite{she} about the possible  monadic theories of $\om_2$, that there
are only finitely
many possible local theories of $\om_2$, and that each of them is decidable.
But it seems more
plausible that the situation is much more complicated, as it is shown to be the case for monadic theories of $\om_2$ in \cite{gms}: 
there are in fact  continuum many possible monadic theories of $\om_2$ (in different universes of set theory); 
moreover, for every set of positive integers 
$S \subseteq \om$,  there is a monadic theory of $\om_2$, in some world, which is as complex as $S$.
\nl We then extend the above results by proving that 
for all integers $n, p \geq 1$, 
if $n < p$ then the local theory of $\om_n$ 
is recursive in the local theory of $\om_p$ and also in the local theory of $\alpha$ where 
$\alpha$ is any ordinal of cofinality $\om_n$. 
\nl Some of these  new results are seemingly far from problems arising in concrete
applications
studied in computer science. However our main result is  obtained by
encoding (Kurepa) trees in models of a local sentence and
methods used here for such coding
might be very useful for problems arising in computer science
where (finite or infinite) trees are a widely used tool.

 The paper is organized as follows. In section 2 we review some previous definitions and 
results about local sentences. In section 3 we prove  new decidability results. 
Our main result on the  local theory of $\om_2$ is proved  in section 4.   
Some results on the local theories of $\om_n$, $n\geq 1$, are stated in section 5.

\section{Review of previous results}

In this paper the (first order) signatures are finite, always contain one
binary
predicate symbol $=$ for equality, and can contain both functional
and relational symbols.

\noi When $M$ is a structure in a signature $\Lambda$, $|M|$ is the domain of $M$. 
\nl If $f$ is  a function symbol (respectively, $R$ is a  relation symbol, $a$ is a 
 constant symbol) in $\Lambda$, then  
 $f^M$ (respectively, $R^M$, $a^M$) is the interpretation in the structure 
$M$ of  $f$ (respectively, $R$, $a$). 
\nl Notice that, when the meaning is clear, the superscript $M$ in 
$f^M$, $R^M$, $a^M$, will be sometimes omitted in order to simplify the presentation. 

 For a structure $M$  in a signature $\Lambda$ 
 and $X\subseteq |M|$,
we define:
\nl $cl^1(X, M)=X \cup \bigcup_{\{f ~{\rm n-ary~ function~ of ~} \Lambda
~\} } ~f^M(X^n)
\cup \bigcup_{\{a ~{\rm ~ constant~ of ~} \Lambda ~\} } a^M $
\nl $cl^{n+1}(X, M)=cl^1(cl^n(X, M), M) \quad {\rm ~ for ~an ~integer~}
n\geq 1$
\nl and $cl(X, M)=\bigcup_{n\geq 1} cl^n(X, M)$ is the closure of $X$ in
$M$.

 The
signature of a first order sentence $\varphi$, i.e. the set of
non logical symbols appearing in $\varphi$, is denoted S($\varphi$).  As usual 
$M\models\varphi$ means that the sentence $\varphi$ is satisfied in the structure $M$, i.e. 
that $M$ is a model of $\varphi$.

\begin{Deff}\label{defloc} A first order sentence $\varphi$
is local if and only if:
\begin{enumerate}
\item[(a)] $M\models\varphi$ and $X\subseteq|M|$ imply
$cl(X,M)\models\varphi$
\item[(b)] $\exists n\in \mathbb{N}$ such that $\forall M$, if $ M\models
\varphi$ and
$X \subseteq |M|$,
then $ cl(X,M)=cl^n(X, M)$, (closure in models of $\varphi$ takes less than
$n$ steps).

\end{enumerate}
\end{Deff}

\noi For a local sentence $\varphi$, $n_\varphi$ is the
smallest integer $n\geq 1$ satisfying $(b)$ of the above definition.
In this definition, $(a)$ implies that a local sentence
$\varphi$ is always equivalent to a universal
sentence, so we may assume that this is always the case. 

\begin{Exa}
Let $\varphi$ be the sentence in the signature $S(\varphi)=\{<,
P, i, a\}$,
where $<$ is a binary relation symbol, $P$ is a unary relation symbol, $i$
is a unary
function symbol, and $a$ is a constant symbol, which is the conjunction of:

\begin{enumerate}
\ite[(1)] $\fa xyz [ (x\leq y \vee y\leq x) \wedge ((x\leq y \wedge y\leq
x) \lra x=y) \wedge
((x\leq y \wedge y\leq z) \ra x\leq z) ]$,
\ite[(2)] $\fa x y [ ( P(x) \wedge \neg P(y) ) \ra x<y ]$,
\ite[(3)] $\fa x y [ ( P(x) \ra i(x)=x ) \wedge ( \neg P(y) \ra
P(i(y)) ) ]$,
\ite[(4)] $\fa x y [ ( \neg P(x) \wedge \neg P(y) \wedge x \neq y ) \ra
i(x) \neq i(y) ]$,
\ite[(5)] $\neg P(a)$.
\end{enumerate}

\noi We now explain the meaning of the above sentences $(1)$-$(5)$.
\nl Assume that $M$ is a model of $\vp$.
The sentence $(1)$ expresses that $<$ is interpreted in $M$ by a linear
order; $(2)$ expresses
that $P^M$ is an initial segment of the model $M$; $(3)$ expresses that the
function $i^M$
is trivially defined by $i^M(x)=x$ on $P^M$ and is defined from $\neg P^M$
into $P^M$.
$(4)$ says that $i^M$ is an injection from $\neg P^M$ into $P^M$ and $(5)$
ensures that
the element $a^M$ is in $\neg P^M$.
\nl The sentence $\varphi$ is a conjunction of universal sentences thus it
is
equivalent to a universal one, and closure in its models takes at most
two steps (one adds the constant $a$ in one step then takes the closure
under the function $i$).
Thus $\varphi$ is a local sentence.
\nl If we consider only the order types of {\it well ordered} models of
$\varphi$,
we can easily see that $\varphi$ has a model of order type $\alpha$, for
every finite ordinal
$\alpha \geq 2$ and for every infinite ordinal $\alpha$ which is not a cardinal.

\end{Exa}

Many more
examples of local
sentences will be given later in Sections 4 and 5. The reader may
also find many
other ones in the papers \cite{ress,fr,loc} \cite{of,toploc,cloloc}.

The set of local sentences is recursively enumerable but not recursive
\cite{loc}.
However there exists a ``recursive presentation"
up to logical equivalence
of all local sentences.

\begin{The}[Ressayre, see \cite{loc}]\label{rec}
There exists a recursive set ${\bf L}$ of local sentences and a recursive
function ${\bf F}$
such that:
\begin{enumerate}
\ite[1)] $\psi$ local $\longleftrightarrow \exists \psi'\in {\bf L}$ such that
$\psi \equiv \psi'$.
\ite[2)] $\psi'\in {\bf L} \longrightarrow n_{\psi'}={\bf F}(\psi')$.
\end{enumerate}
\end{The}

\noi The elements of ${\bf L}$ are the $\psi \wedge C_n$, where $\psi$ run
over
the universal formulas and $C_n$ run over the universal formulas in the
signature $S(\psi)$
which express that closure in a model takes at most $n$ steps.
\nl $\psi \wedge C_n$ is local and $n_{\psi \wedge C_n} \leq n$. Then we can
compute
$n_{\psi \wedge C_n}$, considering only finite models of cardinal $\leq m$,
where $m$
is an integer depending on $n$. And each local sentence $\psi$ is equivalent
to
a universal formula $\theta$, hence $\psi \equiv \theta \wedge C_{n_\psi}$.

   From now on we shall assume that the signature of local sentences
contain
a binary predicate $<$ which is interpreted by a linear ordering in all
of their models.
\nl We recall now the stretching theorem for local sentences.
Below, semi-monotonic, special,
and monotonic indiscernibles are particular kinds of indiscernibles
which are precisely defined in \cite{fr}.

\begin{The}[\cite{fr}]\label{stretching}
For each local sentence $\varphi$ there exists a positive
integer $N_\varphi$ such that
\begin{enumerate}
\ite[(A)] $\varphi$ has arbitrarily large finite models if and only if
$\varphi$ has
an infinite model if and only if $\varphi$ has a finite model generated by
$N_\varphi$
indiscernibles.
\ite[(B)] $\varphi$ has an infinite well ordered model
if and only if $\varphi$ has a finite model generated by $N_\varphi$
semi-monotonic
indiscernibles.
\ite[(C)] $\varphi$ has a model of order type $\om$
if and only if $\varphi$ has a finite model generated by $N_\varphi$ special
indiscernibles.
\ite[(D)] $\varphi$ has well ordered models of unbounded order types in
the ordinals
if and only if $\varphi$ has a finite model generated by $N_\varphi$
monotonic
indiscernibles.

\end{enumerate}
\noi To every local sentence $\varphi$ and every ordinal $\alpha$ such that
$\om \leq \alpha <\om^\om$ one can associate by an effective procedure a
local sentence
$\vp_\alpha$, a unary predicate symbol $P$ being in the signature
S($\varphi_\alpha$),
such that:
\begin{enumerate}
\ite[($C_\alpha$)] $\varphi$ has a well ordered model of order type
$\alpha$
if and only if $\varphi_\alpha$ has a finite model $M$ generated by
$N_{\varphi_\alpha}$
semi-monotonic indiscernibles into $P^M$.
\end{enumerate}

\end{The}

\noi
The integer $N_\varphi$ can be effectively computed from $n_\varphi$
and $q$ where $\varphi=\fa x_1 \ldots \fa x_q \nl \theta(x_1, \ldots, x_q)$ and
$\theta$ is an
open formula, i.e. a formula without quantifiers.
If $v(\varphi)$ is the maximum number of variables of terms of complexity
$\leq n_\varphi +1$
(resulting by at most $n_\varphi +1$ applications of function symbols)
and $v'(\varphi)$ is the maximum number of variables of an atomic formula
involving terms of complexity $\leq n_\varphi +1$ then
$N_\varphi = max \{ 3v(\varphi) ; v'(\varphi) + v(\varphi) ; q.v'(\varphi)
\}$.

   From Theorem \ref{stretching} we can prove the decidability of several
problems about
local sentences. For instance $(C)$ states that a local sentence
$\varphi$ has an infinite well ordered model
iff it has a {\it finite } model generated by $N_\vp$ semi-monotonic
indiscernibles.
Therefore in order to check the existence of an infinite well ordered
model of $\vp$
one can only consider models whose cardinals are bounded by an integer
depending
on $n_\vp$ and $N_\vp$, because
closure in models of $\vp$ takes at most $n_\vp$ steps. This can be done in
a finite
amount of time.

   Notice that the set of local sentences is not recursive
so the algorithms given by the following theorem are applied
to local sentences in the recursive set ${\bf L}$ given by Proposition
\ref{rec}.
In particular $\vp$ is given with the integer $n_\vp$.

\begin{The}[\cite{fr}]\label{decloc} It is decidable,
for a given local sentence $\varphi$, whether
\begin{enumerate}
\ite[(1)] $\varphi$ has arbitrarily large finite models.
\ite[(2)] $\varphi$ has an infinite model.
\ite[(3)] $\varphi$ has an infinite well ordered model.
\ite[(4)] $\varphi$ has well ordered models of unbounded order types in
the ordinals.
\ite[(5)] $\varphi$ has a model of order type $\alpha$,
where $\alpha <\om^\om$ is a given ordinal.

\end{enumerate}

\end{The}

\noi These decidable problems $(1)-(4)$ and $(5)$ (at least for
$\alpha=\om$)
are in the class ${\bf NTIME(2^{O(n.log(n))})}$, (and even probably of lower
complexity):
\nl Using non determinism a Turing machine may guess a finite structure $M$
of signature
S($\varphi$) generated by $N_\vp$ elements $y_1, \ldots y_{N_\vp}$
in at most $n_\vp$ steps. Then, assuming
$\varphi=\fa x_1 \ldots \fa x_q \theta(x_1, \ldots, x_q)$ where $\theta$ is
an
open formula, the Turing machine checks that $\theta(x_1, \ldots, x_q)$
holds
for all $x_1 \ldots x_q$ in $M$,
and that the elements $y_1, \ldots y_{N_\vp}$ are indiscernibles
(respectively,
semi-monotonic, special, monotonic, indiscernibles) in $M$.
\nl On the other hand B\"uchi's procedure to decide whether a monadic
second order
formula of size $n$ of $S1S$ is true in the structure $(\om, <)$ might run
in time
${\bf \underbrace{2^{2^{.^{.^{2^n}}}}}_{O(n)}}$, \cite{bu62,saf}.
Moreover Meyer proved that one cannot essentially improve this result:
the monadic second order theory of
$(\om, <)$ is not elementary recursive, \cite{mey}.
\nl We know that the expressive power of local sentences is much greater
than that
of monadic second order sentences hence this is a
remarkable fact that decision
algorithms for local sentences given by Theorem \ref{decloc} are
of much lower complexity than the algorithm for decidability of the monadic
second order
theory $S1S$ of one successor over the integers.
\nl Notice however that    the nonemptiness problem for B\"uchi automata  is known to be logspace-complete for  the complexity 
class $\bf{NLOGSPACE}$ which is included in the class $\bf{DTIME(Pol)}$ of  problems 
which can be solved in deterministic polynomial time \cite{vw,bgg}.  Moreover there is a linear time algorithm for deciding the 
 nonemptiness problem for B\"uchi automata which is nowadays  very useful for many applications in the domain  of  specification and verification 
of non terminating systems, see for example \cite{bv}.  

\section{More decidability results}\label{section3}

\noi We assume in this section that the function symbols
of a local sentence $\vp$ are at most unary.
We shall prove in this case some
more decidability results which rely on an extension of the stretching
Theorem \ref{stretching}.

   The cardinal of a set $X$ will be denoted by $card(X)$.
\nl We recall that the infinite cardinals are usually denoted by
$\aleph_0, \aleph_1, \aleph_2, \ldots , \aleph_\alpha, \ldots$
The cardinal $\aleph_\alpha$ is also denoted by $\om_\alpha$,
as usual when it is considered as an ordinal.

We recall now the notions of cofinality 
of an ordinal and of regular cardinal which may be found for instance 
in \cite{dev,jech}. 
\nl Let $\alpha$ be a limit ordinal, the cofinality of $\alpha$, denoted $cof(\alpha)$, 
is the least ordinal $\beta$ such that there exists a strictly increasing sequence of ordinals 
$(\alpha_i)_{i<\beta}$, of length $\beta$, such that 
$$\fa i< \beta  ~~~~~ \alpha_i < \alpha ~~~ \mbox{ and } $$
$$\sup_{i< \beta} \alpha_i = \alpha$$
\noi This definition is usually extended to 0 and to the successor ordinals:
$$cof(0)=0 \mbox{ and } cof(\alpha +1)=1 \mbox{ for every ordinal  } \alpha.$$

The cofinality of a limit ordinal is always a limit ordinal satisfying:
$$\om \leq cof(\alpha) \leq \alpha $$
\noi $cof(\alpha)$ is in fact a cardinal.  
 A cardinal $k$ is said to be {\it regular} iff $cof(k)=k$. Otherwise $cof(k)<k$ and 
the cardinal $k$ is said to be {\it singular}. 

We recall now the notion of special indiscernibles, \cite{fr},
in that particular case where all function symbols hence all terms of
$S(\vp)$ are unary.
\nl A set $X$ included in a structure $M$, having a linear ordering $<$ in
its signature,
is a set of indiscernibles iff whenever $\bar{x}$ and $\bar{y}$ are order
isomorphic sequences
from $X$ they satisfy in $M$ the same atomic sentences. The indiscernibles
of $X$ are special
iff they satisfy $(i)$ and $(ii)$:

$(i)$~~ for all $x < y $ in $X$ and all terms $t$: $t(x) < y$.

$(ii)$~~for all $x < y $ in $X$ and all terms $t$: $t(y) < x \ra t(y)=t(z)$
for all elements
$z > x$ of $X$ (i.e. $t$ is constant on $\{z\in X \mid z>x \}$).

\begin{The}\label{unary} For each local sentence $\vp$ whose function symbols are at
most unary, there is a positive integer $N_\vp$ such that, for each regular
cardinal
$\om_\alpha$,
the following statements are equivalent:
\begin{itemize}

\ite[(a)] $\vp$ has an $\om$-model.

\ite[(b)] $\vp$ has a finite model generated by $N_\vp$ special
indiscernibles.

\ite[(c)] $\vp$ has a $\beta$-model, for all limit ordinals $\beta$.

\ite[(d)] $\vp$ has an $\om_\alpha$-model.

\end{itemize}
\end{The}

\proo It is proved in \cite{fr} that for each local sentence $\vp$ there
is a positive
integer $N_\vp$ such that $(a)$ is equivalent to $(b)$.
\nl To prove $(a) \ra (c)$ assume that $\vp$ has an $\om$-model $M$. Then
it is proved
in \cite{fr} that there exists an infinite set $X$ of special indiscernibles
in $M$.
Recall that every linear order $Y$ can be extended to a model $M(Y)$ of
$\vp$,
called the stretching of $M$ along $Y$, so that:

$(1)$~~$M(X)$ is the submodel of $M$ generated by the set $X$.

$(2)$~~$ Y \subseteq Z$ implies $M(Y) \subseteq M(Z)$.

$(3)$~~Every order embedding $f: Y \ra Z$ has an extension $M(f)$ which is
an embedding of
$M(Y)$ into $M(Z)$.

 Let then $\beta$ be a limit ordinal and $M(\beta)$ be the stretching of
$M$ along $\beta$.
We are going to show that $M(\beta)$ is of order type $\beta$.
The model $M(\beta)$ is generated by the set $\beta$ in a finite number of
steps so
there is a finite set $T_\vp$ of (unary) terms
of the signature $S(\vp)$ such that the domain of $M(\beta)$ is
$\beta \cup \cup_{t\in T_\vp} \cup_{\gamma < \beta } t(\gamma)$. The
indiscernibles are
special thus for each term $t\in T_\vp$, either $t$ is constant on $\beta$
or for all
indiscernibles $x < y < z$ in $ \beta$ we have $x < t(y) < z$. It is then
easy to see
that $M(\beta)$ is of order type $\beta$.
\nl $(c) \ra (d)$ is trivial so it remains to prove $(d) \ra (a)$.
\nl We assume that $\alpha$ is an ordinal and that $M$ is a model of $\vp$
of
order type $\om_\alpha$ where $\om_\alpha$ is a {\it regular} cardinal.
We are going to show that there exists in $M$ an infinite set
of special indiscernibles.
These indiscernibles have to satisfy $(i)$ and $(ii)$ only for terms of
complexity
$\leq n_\vp$ because for each term $t$ of complexity greater than $n_\vp$
there will be
another term $t'$ of complexity $\leq n_\vp$ such that $t(x)=t'(x)$ for all
indiscernibles $x$. This finite set of terms of
complexity $\leq n_\vp$ will be denoted by $T=\{t_1, t_2, \ldots , t_N\}$.
\nl Using the fact that $\om_\alpha$ is a {\it regular} cardinal, we
can firstly construct by induction a strictly increasing sequence
$(x_\delta)_{\delta < \om_\alpha}$ of elements of $M$ such that for each
ordinal
$\delta < \om_\alpha$ and each term $t\in T$ it holds that $t( x_\delta ) <
x_{\delta + 1}$.
We denote $X_0 = \{ x_\delta \mid \delta < \om_\alpha\}$;
this set has cardinal $\aleph_\alpha$.
\nl We consider now the three following cases:

   {\bf First case.}
The set $\{ x_\delta \in X_0 - \{x_0\} \mid t_1( x_\delta ) = x_0 \}$
has cardinal $\aleph_\alpha$. Then we denote this set by $X_0^1$.
\nl {\bf Second case.}
The set $\{ x_\delta \in X_0 - \{x_0\} \mid t_1( x_\delta ) < x_0 \}$
has cardinal $\aleph_\alpha$ and the first case does not hold. The initial
segment
$\{x \in M \mid x < x_0\}$ of $M$ has cardinal smaller than $\aleph_\alpha$
thus there is a
subset of $\{ x_\delta \in X_0 - \{x_0\} \mid t_1( x_\delta ) < x_0 \}$
which has cardinal $\aleph_\alpha$ and on which $t_1$ is constant.
Then we denote this set by $X_0^1$.
\nl {\bf Third case.}
The set $\{ x_\delta \in X_0 - \{x_0\} \mid t_1( x_\delta ) > x_0 \}$
has cardinal $\aleph_\alpha$ and the two first cases do not hold. Then we
call this set
$X_0^1$.

   We can repeat now this process, replacing $X_0$ by $X_0^1$ and the term
$t_1$ by the term
$t_2$, so we obtain a new set $X_0^2 \subseteq X_0^1$ having still cardinal
$\aleph_\alpha$.
Next we repeat the process replacing $X_0^1$ by $X_0^2$ and the term $t_2$
by the term
$t_3$, so we obtain a new set $X_0^3 \subseteq X_0^2$ having still cardinal
$\aleph_\alpha$.
\nl After having considered all terms $t_1, t_2, \ldots , t_N$ we have got a
set
$X_0^N \subseteq X_0^{N-1} \subseteq \ldots \subseteq X_0$. We denote
$X_1=X_0^N$.

   Let $x_{\delta_1}$ be the first element of $X_1$. We can repeat all the
above process
replacing $X_0$ by $X_1$ and $x_0$ by $x_{\delta_1}$. This way,
considering successively each of the terms $t_1, t_2, \ldots , t_N$, we
construct new sets $X_1^N \subseteq X_1^{N-1} \subseteq
\ldots \subseteq X_1^1 \subseteq X_1$, each of them having cardinal
$\aleph_\alpha$, and
we set $X_2=X_1^N$.
\nl Assume now that we have applied this process $K$ times for some integer
$K \geq 2$. Then
we have constructed successively some sets $X_1, X_2, \ldots , X_K$ of
cardinal $\aleph_\alpha$.
Let now $x_{\delta_K}$ be the first element of $X_K$. We can repeat the
above process
replacing $X_0$ by $X_K$ and $x_0$ by $x_{\delta_K}$. This way we construct
a new set
$X_{K+1}=X_K^N$ of cardinal $\aleph_\alpha$.
\nl Then we can construct by induction the sets $X_K$ for all integers
$K \geq 1$.
We set $X = \{ x_{\delta_i} \mid 0 \leq i < \om \}$ where for all $i$,
$x_{\delta_i}$ is the first element of $X_i$.

   Let now $X^{[n]}$ be the set of strictly increasing $n$-sequences of
elements of $X$.
Let $\sim$ be the equivalence relation defined on $X^{[v'(\vp)]}$ by: $x
\sim y$ if and only if
$x$ and $y$ satisfy in $M$ the same atomic formulas of complexity $\leq
n_\vp +1$ (i.e.
whose terms are of complexity
$\leq n_\vp +1$).
Applying the Infinite Ramsey Theorem,
we can now get an infinite set $Y \subseteq X$ such that $Y^{[v'(\vp)]}$ is
contained in
a single equivalence class of $\sim$.
\nl $Y$ is a set of indiscernibles in $M$ because if $z$ and $z'$ are two
elements of
$Y^{[n]}$ for $n\geq v'(\vp)$, then they satisfy in $M$ the same atomic
sentences of complexity
$\leq n_\vp +1$ hence of any complexity by Fact 1 of \cite[page 568]{fr}.
\nl By the above construction
of the set $X$, the indiscernibles of $Y$ are special. Thus the submodel
$M(Y)$ of $M$
generated by $Y$ is a model of $\vp$ of order type $\om$.
\ep

Notice that one cannot omit the hypothesis of the {\it regularity} of
the cardinal $\om_\alpha$ in the above theorem. This is due to the fact that there exists 
a local sentence whose function symbols are at most unary and which has some 
well ordered models of order type $\alpha$, 
for every ordinal $\alpha$ which is not a {\it regular} cardinal. Such an example is given in 
\cite{fr}. We are going to recall it now because some steps of its construction 
will be also useful later. 

We recall first the operation $\vp \ra \vp^\star$ over local sentences which was first 
defined by Ressayre in \cite{ress} in order to prove that the class of local languages 
is closed under star operation. 
\nl For each local sentence $\varphi$, the signature of the first order sentence 
 S($\varphi^\star$) is 
  S($\varphi$) to which is added a unary function symbol  
$I$ and in which  every constant symbol $e$ is replaced by a unary function symbol $e(x)$.
\nl $\varphi^\star$ is the sentence defined by the  conjunction of:

\begin{itemize}
\ite[(1)] ( $<$ is a linear order ),

\ite[(2)]  $\fa yz [I(y) \leq y $ and $ (y \leq z \ra I(y) \leq I(z)) $ and 
$ (I(y) \leq z \leq y \ra I(z)=I(y) ) ] $,

\ite[(3)] $\fa xy [I(x)=I(y) \ra e(x)=e(y)] $, for each constant  e of the  signature 
S($\varphi$) of  $\varphi$,

\ite[(4)] $\fa x_1 \ldots x_n [ (\bigvee_{i, j \leq n} I(x_i) \neq I(x_j)) \ra 
f(x_1 \ldots x_n)=min(x_1 \ldots x_n) ]$ , for each  n-ary function $f$  of  S($\varphi$),

\ite[(5)] $\fa x_1 \ldots x_n [ (\bigwedge_{i, j \leq n} I(x_i)=I(x_j) ) 
\ra I(f(x_1 \ldots x_n))=I(x_1)]$,
 for each  n-ary function $f$  of  S($\varphi$),

\ite[(6)] $\fa x \varphi^x $, where  $\varphi^x$  is the local sentence  $\varphi$ 
in which every constant e is replaced by the term e(x) and each  quantifier is relativized
to the set $\{ y \mid I(y)=I(x) \}. $
\end{itemize}

\noi We now explain the meaning of sentences $(1)$-$(6)$. 
Sentence $(2)$ is used to divide a model $M$  of $\varphi^\star$ 
into successive segments.
The function $I^M$ is constant on each of these segments and the image $I^M(x)$
of an element
$x$ is the first element of the segment containing $x$.
 Sentence $(3)$ expresses that each unary function $e^M$ obtained from a constant symbol 
$e\in$ S($\varphi$) is constant on every segment of the model. 
$(4)$ and $(5)$ express that,  for each function symbol $f \in$ S($\varphi$), each segment of 
the model is closed under the function $f^M$ and that $f^M$ is trivially defined by 
$f^M(x_1 \ldots x_n)=min(x_1 \ldots x_n)$ when at least two of the elements $x_i$ belong to 
different segments.  
Finally sentence $(6)$ expresses that each structure which is obtained by 
restricting  some segment of the model  to the signature of $\vp$ is a model of $\vp$. 
 This implies that  models of  $\varphi^\star$   are essentially  direct sums of 
 models of $\varphi$. 
\nl It is easy to see that $n_{\varphi^\star}=n_\vp +1$. Closure in models of $\varphi^\star$   
takes at most $n_\vp +1$ steps: one takes the closure under the function $I$ then the closure 
under functions of  S($\varphi$) in $n_\vp$ steps. 

We recall now the operation $(\vp, \psi) \ra \vp^{\star \psi}$ over local sentences which is an 
extension of the operation $\vp \ra \vp^\star$ and is defined in \cite{fr}. 

We assume that S($\varphi^\star$)$\cap $ S($\psi$)=$\{<\}$. 
Then  S($\varphi^{\star\psi}$)=S($\varphi^\star$) $\cup$ S($\psi$) $\cup$ $\{ P \}$, 
where $P$ is a new unary predicate 
symbol not in  S($\varphi$) $\cup$ S($\psi$). 
\nl $\varphi^{(\star\psi)}$ is the  conjunction of :

\begin{itemize}

\ite[(1)] $\varphi^\star$,

\ite[(2)] $\fa x [P(x) \leftrightarrow I(x)=x],$

\ite[(3)] $\fa x_1 \ldots x_k [(\bigwedge_{i=1}^k P(x_i)) \ra P(t(x_1 \ldots x_k)) ]$,
 for each  k-ary function $t$ of  S($\psi$),

\ite[(4)] $ P(a)$, for each constant $a$  of  S($\psi$),

\ite[(5)] $\fa x_1  \ldots x_n  
[(\bigvee_{i=1}^n \neg P(x_i) )\ra t(x_1 \ldots x_n)=min(x_1 \ldots x_n) ]$,
for each  n-ary function $t$  in $S(\psi$).

\ite[(6)] $\fa x_1 \ldots x_k [Q(x_1 \ldots x_k) \ra P(x_1)\wedge \ldots \wedge P(x_k) ]$, 
for each k-ary predicate symbol $Q$  of S($\psi$)

\ite[(7)] $\fa x_1 \ldots x_n[ (\bigwedge_{i=1}^n P(x_i))\rightarrow \psi_1(x_1 \ldots x_n)]$, 
where 
 $\psi = \fa x_1 \ldots x_n \psi_1(x_1 \ldots x_n)$ and  $\psi_1$ is an open formula, 

\end{itemize}

\noi We now explain the meaning of $(1)$-$(7)$. Sentence $(1)$ is $\varphi^\star$ so it 
expresses that a model $M$ is essentially a direct sum of models of $\vp$ . $(2)$ says 
that in such a model $M$, $P^M$ is the set of first elements of the segments of $M$ defined 
with the function $I^M$. $(3)$-$(5)$ are used to 
ensure that $P^M$ is closed under functions of S($\psi$) and that these functions are 
trivially defined elsewhere. $(6)$ says that for every k-ary predicate $Q$ in  S($\psi$) the 
set $Q^M$ is included into $(P^M)^k$. 
Sentence $(7)$ expresses that  the restriction of $M$ to the set 
$P^M$ and to the signature of $\psi$ is a model of $\psi$. 
\nl It is easy to see that $n_{\varphi^{(\star\psi)}}=n_{\varphi} + n_\psi + 1$;  
to take closure of a set $X$ in a model of $\varphi^{(\star\psi)}$ one takes the closure 
under the function $I$, then under the functions of S($\psi$) in $n_\psi$ steps, then 
under the functions of S($\varphi$) in $n_\vp$ steps. 
\nl The models of  $\varphi^{(\star\psi)}$ essentially are direct sums of  
models of $\varphi$ , these  models being ordered by the  order  type  of a model of  $\psi$.

We are mainly interested in this paper by {\it well ordered} models of local sentences, 
so we now recall the notion of spectrum of a local sentence $\vp$. 
As usual the class of all ordinals is denoted by {\bf On}.

\begin{Deff}
Let $\vp$ be a local sentence; the spectrum of $\vp$ is 
$$Sp(\vp) = \{ \alpha \in {\bf On} \mid \vp \mbox{ has a model of order type } \alpha \}$$
\noi and the infinite spectrum of $\vp$ is 
$$Sp_\infty(\vp) = \{ \alpha \in {\bf On} \mid \alpha \geq \om \mbox{ and } 
\vp \mbox{ has a model of order type } \alpha \}$$
\end{Deff} 

\noi The  spectrum of $\varphi^{(\star\psi)}$  depends on the 
spectra of the local sentences $\vp$ and $\psi$ ans is given by the following proposition. 

\begin{Pro}
Let $\vp$ and $\psi$ be some local sentences, then $\varphi^{(\star\psi)}$  is a local sentence 
and its spectrum is
$$Sp( \varphi^{(\star\psi)} ) = \{ \sum_{\alpha < \nu}{ a_\alpha } \mid \nu \in Sp(\psi) 
\mbox{ and }  \fa \alpha < \nu ~
a_\alpha \in Sp(\vp)  \}$$

\end{Pro}

\noi We can now construct a local sentence which has models of order type $\alpha$ for every 
infinite ordinal $\alpha$ which is not a regular cardinal \cite{fr}. 
\nl Let $\theta$ be a local sentence in the signature S($\theta$)=$\{<, a\}$ 
which just expresses 
that the constant symbol $a$ is interpreted by the last element of a model. 
Then the spectrum of 
$\theta$ is the class of successor ordinals. 
\nl And let $\beta$ be a local sentence in the signature S($\beta$)=$\{<, P, s\}$ where 
$P$ is a unary predicate symbol and $s$ is a unary function symbol, which expresses  that 
in a model $M$, the set $P^M$ is an initial segment of the model and that $s^M$ is a 
strictly non decreasing 
involution from $P^M$ onto $\neg P^M$. Then the spectrum of $\beta$ is the class of ordinals 
of the form $\alpha.2$ for some ordinal $\alpha$. 
\nl It holds by construction that 
there are not in the signature of $\theta^{\star \beta}$  any function symbols of arity 
greater than $1$ and 
we can  verify that $Sp_\infty(\theta^{\star \beta})=\{ \alpha \geq \om \mid 
\alpha \mbox{ is not a regular cardinal } \}$. In particular 
the sentence $\theta^{\star \beta}$  has a model of order type $\om_\alpha$ for every 
singular cardinal $\om_\alpha$ but it has no model of order type $\om$. So the hypothesis 
of the regularity of the cardinal $\om_\alpha$ was necessary in Theorem \ref{unary}. 

Return now to decision algorithms given by stretching theorems. 
By Theorem \ref{decloc} it is decidable whether a local sentence $\vp$
has
an $\om$-model so Theorem \ref{unary} implies also  the following decidability result.

\begin{The}\label{decloc2} It is decidable,
for a given local sentence $\varphi$ whose function symbols are at
most unary, and a given regular cardinal $\om_\alpha$, whether:
\begin{enumerate}
\ite[(1)] $\vp$ has an $\om_\alpha$-model
\ite[(2)] $\vp$ has a $\beta$-model for all
limit ordinals $\beta$.

\end{enumerate}

\end{The}

\noi So in particular one can decide,
for a given local sentence $\varphi$ whose function symbols are at
most unary, whether $\varphi$ has a model of order type $\om_1$,
(respectively,
$\om_2$, $\om_n$ where $n$ is a positive integer).

As mentioned in the introduction 
it is interesting to know that the class  $ LOCAL(1)$
of local sentences with at most unary function symbols 
has more decidability properties 
because it has already a great expressive power.
\nl In particular 
$ LOCAL(1)$ can define all \orl s \cite{loc}, but also some ${\bf
\Si^0_{n}}$-complete and
some ${\bf \Pi^0_{n}}$-complete Borel sets for every integer $n\geq
1$,
\cite{toploc}.

Moreover it is easy to see that local $\om$-languages 
satisfy an extension of B\"uchi's lemma. 
Recall that this lemma states that a regular $\om$-language is non-empty if and 
only if it contains an ultimately periodic $\om$-word, i.e. an $\om$-word in the form 
$u.v^\om$ for some {\it finite } words $u$ and $v$. 
\nl On the other hand by the proof of 
the Stretching Theorem \ref{stretching} (C) we know that a local 
$\om$-language $L(\vp) \subseteq \Sio$ 
is non-empty if and only if it contains an $\om$-word which is the reduction 
to  the signature $\Lambda_\Si = \{<, (P_a)_{a\in \Sigma} \}$ (of words over $\Si$) 
of an $\om$-model of $\vp$ 
generated by special indiscernibles. 
\nl If the function symbols of the 
 local sentence $\vp$ are at most unary then it is easy to see 
that such a reduction of an $\om$-model of $\vp$ 
generated by special indiscernibles is always an  ultimately periodic $\om$-word.

\section{The local theory of $\omega_2$}

\noi It was proved in \cite{fr} that there exists a local sentence $\psi$
(whose signature
contains binary function symbols) having well ordered
models of order type $\alpha$ for every ordinal $\alpha$
in the segment $[\om ; 2^{\aleph_0}]$ but not
any well ordered model of order type $\alpha$ for $card(\alpha) >
2^{\aleph_0}$.
On the other hand it is well known that the continuum hypothesis $CH$ is
independent of
the axiomatic system $ZFC$. This means that there are some models of $ZFC$
in which
$ 2^{\aleph_0}=\aleph_1$ and some others in which $ 2^{\aleph_0}\geq
\aleph_2$.
Therefore the statement ``$\psi$ has a model of order type $\om_2$" is
independent
of $ZFC$.
\nl However if we assume the continuum hypothesis and even the generalized
continuum hypothesis
$GCH$ saying that, for every cardinal $\aleph_\alpha$,
$2^{\aleph_\alpha}=\aleph_{\alpha+1}$,
then the above result of \cite{fr} does not imply a similar independence
result.

   Nevertheless we are going to prove the existence of a local sentence
$\Phi$
such that ``$\Phi$ has a model of order type $\om_2$" is independent
of $ZFC+GCH$.
\nl For that purpose we shall use results about Kurepa trees which we now
recall.

   A partially ordered set $( T, \prec_T )$ is called a tree if for every
$t\in T$ the set
$\{s\in T \mid s \prec_T t \}$ is well ordered under $\prec_T$. Then the
order type of the set
$\{s\in T \mid s \prec_T t \}$ is called the height of $t$ in $T$ and is
denoted by $ht(t)$.
We shall not distinguish a tree from its base set.
\nl For every ordinal $\alpha$ the $\alpha$-th level of $T$ is
$T_\alpha = \{ t\in T \mid ht(t)=\alpha \}$.
\nl The height of $T$, denoted by $ht(T)$, is the smallest ordinal $\alpha$
such that
$T_\alpha = \emptyset$.
\nl A branch of $T$ will be a linearly ordered subset of $T$ intersecting
every non-empty
level of $T$. The set of all branches of $T$ will be denoted
$\mathcal{B}(T)$.
\nl A tree $T$ is called an $\om_1$-tree if $card(T)=\aleph_1$ and
$ht(T)=\om_1$. An $\om_1$-tree
$T$ is called a Kurepa tree if $card(\mathcal{B}(T))>\aleph_1$
and for every ordinal $\alpha < \om_1$,
$card(T_\alpha) < \aleph_1$.

   Recall now the well known results about Kurepa trees, \cite{dev}:

\begin{The}\label{theo1}
\noi
\begin{enumerate}
\ite If $ZF$ is consistent so too is the theory:
$ZFC + GCH + ``\mbox{ there is a Kurepa tree }"$.
\ite If the theory
$ZFC + ``\mbox{ there is an inaccessible cardinal }"$
is consistent so too is the theory
$ZFC + GCH + ``\mbox{ there are no Kurepa trees }"$.
\ite If the theory
$ZFC + ``\mbox{ there are no Kurepa trees }"$
is consistent so too is the theory
$ZFC + ``\mbox{ there is an inaccessible cardinal }".$
\end{enumerate}
\end{The}

\noi In order to use the above result in the context of local sentences we
state now
the main technical result of this section.

\begin{The}\label{Phi}
There exists a local sentence $\Phi$ such that:
$$[ \Phi \mbox{ has an } \om_2\mbox{-model }] ~~ \longleftrightarrow
~~\mbox{
[ there is a Kurepa tree ]}.$$
\end{The}

\noi To prove this theorem we shall firstly state the two following lemmas.

\begin{Lem}\label{phi0}
There exists a local sentence $\vp_0$ such that $\vp_0$ has a well ordered
model of order
type $\om$ but has no well ordered model of order type $> \om$.
\end{Lem}

\proo Such a sentence is given in \cite{fr} in the signature
$S(\vp_0)=\{<, P, f, p_1, p_2 \}$, where $P$ is a unary predicate, $f$ is a
binary function,
and $p_1, p_2$ are unary functions.
\ep

\begin{Lem}\label{phi1}
There exists a local sentence $\vp_1$ such that $\vp_1$ has well ordered
models of order
type $\alpha$, for every ordinal $\alpha \in [\om, \om_1]$,
but has no well ordered model of order type $> \om_1$.
\end{Lem}

\proo We give below the sentence $\vp_1$ in the signature
$S(\vp_1)=S(\vp_0) \cup \{Q, g \}=\{<, P, f, p_1, p_2, Q, g \}$,
where $Q$ is a unary predicate and $g$ is a binary function.
$\vp_1$ is the conjunction of the following sentences $(1)$-$(10)$ whose
meaning is
explained below:

\begin{enumerate}
\ite[(1)] $\fa xyz [ (x\leq y \vee y\leq x) \wedge ((x\leq y \wedge y\leq
x) \lra x=y) \wedge
((x\leq y \wedge y\leq z) \ra x\leq z) ]$,
\ite[(2)] $\fa x y [ ( Q(x) \wedge \neg Q(y) ) \ra x<y ]$,
\ite[(3)] $\fa x y [ ( Q(x) \wedge Q(y) ) \ra f(x, y) \in Q ]$,
\ite[(4)] $\fa x [ Q(x) \ra Q( p_i(x) ) ]$, for each $i\in[1, 2]$,
\ite[(5)] $\fa x y [ ( \neg Q(x) \vee \neg Q(y) ) \ra f(x, y)=x]$,
\ite[(6)] $\fa x [\neg Q(x) \ra p_i(x)= x ]$,
for each $i\in[1, 2]$,
\ite[(7)] $\fa x_1 \ldots x_j \in Q [\vp'_0 (x_1,\ldots , x_j) ]$,
where $\vp_0 = \fa x_1 \ldots x_j \vp'_0 (x_1,\ldots , x_j)$ with $\vp'_0$
an open formula,
\ite[(8)] $\fa x y [ ( \neg Q(x) \wedge \neg Q(y) \wedge y<x ) \ra Q( g(x,
y) ) ]$,
\ite[(9)] $\fa x y z [ ( \neg Q(x) \wedge \neg Q(y) \wedge \neg Q(z) \wedge
y<z<x ) \ra
g(x, y)\neq g(x, z) ]$,
\ite[(10)] $\fa x y [ ( Q(x) \vee Q(y) \vee \neg (y<x) ) \ra g(x,
y)=x ]$.

\end{enumerate}

\noi We now explain the meaning of the above sentences $(1)$-$(10)$.
\nl Assume that $M$ is a model of $\vp_1$.
The sentence $(1)$ expresses that $<$ is interpreted in $M$ by a linear
order; $(2)$ expresses
that $Q^M$ is an initial segment of the model $M$; $(3)$ and $(4)$ state
that $Q^M$ is closed
under the functions of $S(\vp_0)$ while $(5)$ and $(6)$ state that these
functions are trivially defined
elsewhere; $(7)$ means that the restriction of the model $M$ to the domain
$Q^M$ and
to the signature of $S(\vp_0)$ is
a model of $\vp_0$; Finally $(8)$ and $(9)$ ensure that, for each $x\in \neg
Q$,
the binary function $g$ realizes an injection from the segment $\{ y \in
\neg Q \mid y < x \}$
into $Q$ and $(10)$ states that the function $g$ is trivially defined where
it is not useful
for that purpose.
\nl The sentence $\vp_1$
is a conjunction of universal sentences thus it is
equivalent to a universal one, and closure in its models takes at most
$n_\vp +1$ steps: one applies first the function $g$ and then the functions
of $S(\vp_0)$.
Thus the sentence $\vp_1$ is local.
\nl Consider now a well ordered model $M$ of $\vp_1$. The restriction of
$M$ to the domain $Q^M$ and to the signature of $S(\vp_0)$ is
a well ordered model of $\vp_0$ hence it is of order type $\leq \om$. But
the function $g$
defines an injection from each initial
segment of $\neg Q$ into $Q$ thus each initial segment of $\neg Q$ is
countable and this implies
that the order type of $\neg Q^M$ is smaller than or equal to $\om_1$.
Finally we have proved
that the order type of $M$ is $\leq \om_1$.
\nl Conversely it is easy to see that every ordinal $\alpha \in [\om,
\om_1]$ is the order type
of some model of $\vp_1$.
\ep

   Return now to the construction of the sentence $\Phi$ given by Theorem
\ref{Phi}.
We are going to explain this construction by several successive steps.

   A model $M$ of $\Phi$ will be totally ordered by $<$ and will be
the disjoint union of four successive segments. This will
be expressed by the following sentence $\Phi_1$ in the signature
$S(\Phi_1)=\{P_0, P_1, P_2, P_3\}$, where $P_0$, $P_1$, $P_2$, $P_3$,
are unary predicate symbols. $\Phi_1$ is the conjunction of:

\begin{itemize}
\ite[(1)] $\fa xyz [ (x\leq y \vee y\leq x) \wedge ((x\leq y \wedge
y\leq x) \lra x=y) \wedge
((x\leq y \wedge y\leq z) \ra x\leq z) ]$,
\ite[(2)] $\fa xy \bigwedge_{0\leq i<j\leq 3} [ ( P_i(x) \wedge P_j(y) )
\ra
x < y ]$.
\end{itemize}

\noi We want now to ensure
that, if $M$ is a well ordered model of $\Phi$, then $P_0^M$ is of order
type $\leq \om$ and
$P_1^M$ is of order type $\leq \om_1$.
For that purpose, the signature of $\Phi$ will contain the signature
$S(\vp_1)=S(\vp_0) \cup \{Q, g \}=\{<, P, f, p_1, p_2, Q, g \}$ and
$\Phi$ will express that
if $M$ is a model
of $\Phi$, then $P_0^M=Q^M$ and the restriction of the model $M$ to
$(P_0^M \cup P_1^M)$ and to the signature of $\vp_1$ is a model of
$\vp_1$.
This is expressed by the following sentence $\Phi_2$ which is the
conjunction of:

\begin{itemize}
\ite[(1)] $\fa x [ Q(x) \leftrightarrow P_0(x) ]$,
\ite[(2)] $\fa x y [ ( x \in P_0\cup P_1 \wedge y \in P_0\cup P_1 )
\ra f(x, y) \in P_0\cup P_1 ]$,
\ite[(3)] $\fa x y [ ( x \in P_0\cup P_1 \wedge y \in P_0\cup P_1 )
\ra g(x, y) \in P_0\cup P_1 ]$,
\ite[(4)] $\fa x [ ( x \in P_0\cup P_1 ) \ra
p_i(x) \in P_0\cup P_1 ]$, for each $i\in[1, 2]$,
\ite[(5)] $\fa x y [ ( x \notin P_0 \cup P_1 \vee y \notin P_0 \cup
P_1 )
\ra f(x, y)=x]$,
\ite[(6)] $\fa x y [ ( x \notin P_0 \cup P_1 \vee y \notin P_0 \cup
P_1 )
\ra g(x, y)=x]$,
\ite[(7)] $\fa x [ x \notin P_0\cup P_1 \ra p_i(x)= x ]$,
for each $i\in[1, 2]$,
\ite[(8)] $\fa x_1 \ldots x_k \in (P_0\cup P_1) [\vp'_1 (x_1,\ldots ,
x_k) ]$,
where $\vp_1 = \fa x_1 \ldots x_k \vp'_1 (x_1,\ldots , x_k)$ with $\vp'_1$
an open formula.

\end{itemize}

\noi Above sentences $(2)$-$(4)$ state that in a model $M$ the set
$(P_0\cup P_1)^M$ is closed
under the functions of $S(\vp_1)$ while $(5)$-$(7)$ state that
these functions are trivially defined
elsewhere; $(8)$ means that the restriction of the model $M$ to the domain
$(P_0\cup P_1)^M$
and to the signature of $S(\vp_1)$ is
a model of $\vp_1$.

   We want now that, in a model $M$ of $\Phi$, the set $P_2^M$ represents
the base
set of a tree $(T, \prec)$. We shall use a binary relation symbol $\prec$.
The following
sentence $\Phi_3$ is the conjunction of:

\begin{itemize}
\ite[(1)] $\fa xy [ x\prec y \ra P_2(x) \wedge P_2(y) ]$,
\ite[(2)] $\fa xyz [ ((x\preccurlyeq y \wedge y\preccurlyeq x) \lra x=y)
\wedge
((x\prec y \wedge y\prec z) \ra x\prec z) ]$.
\ite[(3)] $\fa xy [ x\prec y \ra x < y ]$.
\end{itemize}

\noi Above sentences $(1)$-$(2)$ express that $\prec$ is a partial order on
$P_2$ and the
sentence $(3)$ ensures that, in a well ordered (for $<$ ) model $M$ of
$\Phi_3$,
for every $t\in P_2$,
the set $\{s\in P_2 \mid s \prec t \}$ is well ordered under $\prec$ because
$M$ itself
is well ordered under $<$.

   Moreover we want now that in an $\om_2$-model $M$ of $\Phi$,
the set $P_2^M$ represents the base set of an $\om_1$-tree $T$ whose levels
are countable.
\nl We have firstly to distinguish the different levels of the tree $T$. We
shall use for that
purpose unary functions $I$ and $p$
and the following sentence $\Phi_4$ conjunction of:

\begin{itemize}
\ite[(1)] $\forall xy \in P_2 [(I(y)\leq y) \wedge (y\leq x\rightarrow
I(y)\leq I(x))
\wedge (I(y)\leq x\leq y\rightarrow I(x)=I(y))]$.
\ite[(2)] $\forall xy \in P_2 [ x \prec y \ra I(x) < I(y) ]$,
\ite[(3)] $\forall xyz \in P_2 [ ( x \prec y \wedge z \prec y \wedge I(x) =
I(z) )
\ra x = z ]$,
\ite[(4)] $\forall xy \in P_2 [ I(x) < I(y) \ra ( I( p(I(x), y ) = I(x)
\wedge
p(I(x), y ) \prec y )]$,
\ite[(5)] $\forall xy [ ( \neg P_2(x) \vee \neg P_2(y) \vee I(x)\neq x
\vee I(x) \geq I(y) ) \ra p(x, y )=x ]$,
\ite[(6)] $\forall x [ \neg P_2(x) \ra I(x)=x ]$.

\end{itemize}

\noi Above the sentence $(1)$ is used to divide the segment $P_2$ of
a model of $\Phi_4$ into successive segments.
The function $I$ is constant on each of these segments and the image $I(x)$
of an element
$x\in P_2$ is the first element of the segment containing $x$.
\nl Sentences $(2)$-$(3)$ ensure that if $y \in P_2$ then every element
$x\in P_2$ such that
$ x \prec y$ belongs to some segment
$I_z=\{ w\in P_2 \mid I(w)=I(z)\}$ for some $z < I(y)$.
Moreover for each $z < I(y)$, the segment $I_z$
contains at most one element of
$\{x\in P_2 \mid x \prec y \}$.
\nl The function $p$ is used to ensure that, for each $z < I(y)$,
the segment $I_z$
contains in fact exactly
one element $x\in P_2$ such that $ x \prec y$: the element $p(I(z), y)$.
This is implied by
the sentence $(4)$.
\nl Thus $\Phi_4$ will imply that each segment $I_z$ is really a level
of the tree $T$.
\nl If $y\in P_2$ is at level $\alpha$ of the tree
$T$ and if $x \in P_2$ and $I_x$
represents the $\beta$-th level $T_\beta$ of the tree $T$ for some $\beta <
\alpha$ (so
$I(x) < I(y)$), then
the element $p(I(x), y )$ is the unique element $t \in T_\beta$
such that $t \prec y$.
\nl Finally sentences $(5)$-$(6)$ are used to trivially define the
functions
$p$ and $I$ where they are
not useful as explained above.

   The following sentence $\Phi_5$ will imply that all levels of the tree
$T$ are
countable and that $ht(T)\leq \om_1$ hence also $card(T)\leq \aleph_1$. The
signature of
$\Phi_5$ is $\{ <, P_0, P_1, P_2, I, i, j \}$,
where $i$ and $j$ are two new unary function symbols, and
$\Phi_5$ is the conjunction of:

\begin{itemize}

\ite[(1)] $\forall x [ P_2(x) \ra P_0(i(x)) ]$,
\ite[(2)] $\forall x y [ ( P_2(x) \wedge P_2(y) \wedge I(x)=I(y) \wedge
x\neq y ) \ra
i(x) \neq i(y) ]$,

\ite[(3)] $\forall x [ P_2(x) \ra P_1(j(x)) ]$,

\ite[(4)] $\forall x y [ ( P_2(x) \wedge P_2(y) \wedge x < y ) \ra j(x) <
j(y) ]$,
\ite[(5)] $\forall x [ \neg P_2(x) \ra i(x)=x ]$,
\ite[(6)] $\forall x [ \neg P_2(x) \ra j(x)=x ]$.

\end{itemize}

\noi Above sentences $(1)$-$(2)$ say that the function $i$ is defined from
$P_2$ into $P_0$
and that it is an injection from any level of the tree $T$ into $P_0$. We
have seen that in a
well ordered model $M$ of $\Phi$ the set $P_0^M$ will be of order type $\leq
\om$ thus
each level of the tree will be countable.
\nl Sentences $(3)$-$(4)$ say that the function $j$ is strictly increasing
from $P_2$
into $P_1$ thus in a well ordered model $M$ of $\Phi$ the set
$P_1^M$ hence also $P_2^M$ will be of order type $\leq \om_1$. So we shall
have
$ht(T)\leq \om_1$ and $card(T)\leq \aleph_1$.
\nl Finally sentences $(5)$-$(6)$ are used to trivially define the
functions
$i$ and $j$ on $\neg P_2 = P_0 \cup P_1 \cup P_3$.

   In a well ordered model $M$ of $\Phi$ of order type $\om_2$, the set
$P_2^M$ will
be the base set of an $\om_1$-tree $T$
and the set $P_3^M$ will be identified to a set of branches of $T$.
\nl For that purpose we use two new binary function symbols $h$ and $k$ and
the following
sentence $\Phi_6$, conjunction of:

\begin{itemize}

\ite[(1)] $\forall x y [ ( P_2(x) \wedge P_3(y) ) \ra ( P_2( h(I(x), y) )
\wedge I( h(I(x), y) ) = I(x) ) ]$,

\ite[(2)] $\forall x y z [ ( P_2(x) \wedge P_2(y) \wedge P_3(z) \wedge
I(x)<I(y) ) \ra
h(I(x), z) \prec h(I(y), z) ]$,

\ite[(3)] $\forall x y [ ( \neg P_2(x) \vee \neg P_3(y) \vee x\neq I(x) )
\ra h(x, y)=x ]$,

\ite[(4)] $\forall x y [ ( P_3(x) \wedge P_3(y) \wedge x\neq y ) \ra
( I ( k(x, y) ) = k(x, y) \wedge P_2( k(x, y) ) ) ]$,

\ite[(5)] $\forall x y [ ( P_3(x) \wedge P_3(y) \wedge x\neq y ) \ra
h( k(x, y), x ) \neq h( k(x, y), y ) ) ]$,

\ite[(6)] $\forall x y [ ( \neg P_3(x) \vee \neg P_3(y) \vee x = y )
\ra k(x, y)= x ]$.

\end{itemize}

\noi Above sentences $(1)$-$(2)$ are used to associate a branch $b(z)$ of
$T$ to an element
$z \in P_3$. For each level $T_\alpha$
of the tree which is represented by the segment of $P_2$
whose first element is $I(x)$, the sentence $(1)$ says that $h(I(x), z)$ is
an element
at the same level $T_\alpha$ and $(2)$ says that the elements $h(I(x), z)$,
for
$x\in P_2$, are linearly ordered for $\prec$ hence they form a branch
$b(z)$ of the tree $T$.
\nl The function $k$ is used to associate to two different elements $x$ and
$y$
of $P_3$ a level of the tree $T$, which is represented by the element $k(x,
y)$:
the first element of the segment of $P_2$ representing this level. This is
expressed by the
sentence $(4)$.
\nl The sentence $(5)$ says that, for two distinct elements $x$ and $y$
of $P_3$, the branches $b(x)$ and $b(y)$ differ at the level represented by
$k(x, y)$.
\nl Finally sentences $(3)$ and $(6)$ are used to trivially define the
functions $h$ and $k$
in other cases.

   We have seen that in a well ordered model $M$ of $\Phi$, $P_1^M$ and
$P_2^M$
will be of order type
$\leq \om_1$. The following sentence $\Phi_7$
will then imply that $P_3^M$ is of order type $\leq \om_2$.
Its signature is $\{<, P_1, P_3, l\}$, where $l$ is a binary function
symbol,
and $\Phi_7$ is the conjunction of:

\begin{itemize}

\ite[(1)] $\fa x y [ ( P_3(x) \wedge P_3(y) \wedge y<x ) \ra P_2( l(x,
y) ) ]$,
\ite[(2)] $\fa x y z [ ( P_3(x) \wedge P_3(y) \wedge P_3(z) \wedge y<z<x )
\ra
l(x, y)\neq l(x, z) ]$,
\ite[(3)] $\fa x y [ ( \neg P_3(x) \vee \neg P_3(y) \vee \neg (y<x) ) \ra
l(x, y)=x ]$.

\end{itemize}

\noi Above sentences $(1)$-$(3)$ are in fact very similar to sentences
$(8)$-$(10)$ used
in the construction of the sentence $\vp_1$.
\nl $(1)$ and $(2)$ ensure that, for each $x\in P_3$,
the binary function $l$ realizes an injection from the segment $\{ y \in
P_3 \mid y < x \}$
into $P_2$ and $(3)$ states that the function $l$ is trivially defined where
it is not useful
for that purpose.

   We can now define the sentence
$$\Phi = \bigwedge_{1\leq i \leq 7} \Phi_i$$
\noi in the signature

$$S(\Phi)= \bigwedge_{1\leq i \leq 7} S(\Phi_i)=\{<, P_0, P_1, P_2, P_3, Q,
p_1, p_2, f, g,
\prec, p, I, i, j, h, k, l \}.$$

\noi $\Phi$ is a conjunction of universal sentences thus it is equivalent to
a universal
sentence and closure in its models takes at most $7$ steps: one takes
firstly
closure under the function
$l$ then under the functions $I$ and $k$, then under the functions
$h$ and $p$, then under $i$ and $j$, then under the function $g$, then under
the functions $p_1$
and $p_2$, and finally under the function $f$.
Notice that the two last steps are due to the construction of $\vp_0$
and the fact that $n_{\vp_0}=2$ (see \cite{fr}).
\nl Assume now that $M$ is a well ordered model of $\Phi$. By construction
$P_0^M$ is of
order type $\leq \om$, $P_1^M$ and $P_2^M$ are of order types $\leq \om_1$,
and
$P_2^M$ is the base set of a tree $T$ whose levels are countable. Moreover
every strict
initial segment of $P_3^M$ is of
cardinal $\leq \aleph_1$, so $P_3^M$ is of order type $\leq \om_2$. Finally
we have got that
$M$ itself is of order type $\leq \om_2$.
\nl Suppose now that $M$ is of order type $\om_2$. Then $P_3^M$ also
is of order type $ \om_2$ and for every strict initial segment $J$
of $P_3^M$ there is an injection
from $J$ into $P_2^M$ thus $P_2^M$ is of cardinal $\aleph_1$.
But its order type is $\leq \om_1$,
hence it is in fact equal to $\om_1$.
\nl The tree $T$ is then really an $\om_1$-tree and all its levels are
countable.
Moreover the set $P_3^M$ can be identified to a set of branches of $T$ thus
$card(\mathcal{B}(T))>\aleph_1$ and $T$ is a Kurepa tree.
\nl Conversely if there exists a Kurepa tree, we can easily see
that $\Phi$ has an $\om_2$-model.
\ep

   We can now infer from Theorems \ref{theo1} and \ref{Phi} the following
result
which shows that the local theory of $\om_2$
is not determined by the axiomatic system $ZFC + GCH$.

\begin{The}\label{theo2}
If the theory
$ZFC + ``\mbox{ there is an inaccessible cardinal }"$
is consistent then `` $\Phi$ has an $\om_2$-model " is independent of
$ZFC + GCH $.
\end{The}

\noi Notice that this result can be extended easily to ordinals larger than
$\om_2$.
For instance reasoning as in the construction of the local sentence $\vp_1$
from the local sentence $\vp_0$ (see Lemma \ref{phi1} above),
we can construct by induction, for
each integer $n\geq 2$, a local sentence $\Psi_n$ such that:~ for all
ordinals $\alpha \in
] \om_n, \om_{n+1} ]$, 
\nl ( $\Psi_n$ has an $\alpha$-model ) iff ( $\Phi$ has
an $\om_2$-model )
iff ( there is a Kurepa tree ). This implies the following extension of
Theorem \ref{theo2}.

\begin{The}\label{theo3}
If the theory
$ZFC + ``\mbox{ there is an inaccessible cardinal }"$
is consistent then for each integer $n\geq 2$ and each ordinal $\alpha \in
] \om_n, \om_{n+1} ]$, `` $\Psi_n$ has an $\alpha$-model " is independent of
$ZFC + GCH $.
\end{The}

\noi A similar result can be obtained for larger ordinals 
of cofinality $\om_n$, for an integer $n\geq 2$. 
\nl We can first construct the local sentence $\Theta_2=\theta^{\star \Phi}$, from 
the local sentence $\theta$ given in section \ref{section3} 
whose spectrum 
is the class of successor ordinals,  and the local sentence $\Phi$ we have constructed  
 above.
\nl It is then easy to see that  $\Theta_2$ has not any well ordered model 
whose  order type is  an ordinal $\alpha$ having a cofinality greater than $\om_2$. 
Moerover if $\alpha$ is an ordinal of cofinality $\om_2$ 
then the local sentence $\Theta_2$ has a model of order type $\alpha$ if and only if 
$\Phi$ has an $\om_2$-model. 
\nl In the same way, for each integer $n\geq 2$, 
we can  construct the local sentence $\Theta_{n+1}=\theta^{\star \Psi_n}$, from $\theta$ 
and the local sentence $\Psi_n$ cited 
in the above theorem.
\nl It is then easy to see that  $\Theta_{n+1}$   has not any well ordered 
model whose  order type is an ordinal having a cofinality 
greater than $\om_{n+1}$ because by construction the local sentence 
$\Psi_n$ has no well ordered model of order type greater than $\om_{n+1}$. 
Moreover if $\alpha$ is an ordinal of cofinality $\om_{n+1}$ 
then the local sentence $\Theta_{n+1}$    has a model of order type $\alpha$ 
iff $\Psi_n$ has an $\om_{n+1}$-model iff $\Phi$ has an $\om_2$-model iff there is a Kurepa 
tree. 
\nl So we have got the following extension of Theorem \ref{theo3}.

\begin{The}\label{theo4}
If the theory
$ZFC + ``\mbox{ there is an inaccessible cardinal }"$
is consistent then for each integer $n\geq 2$ and each ordinal $\alpha$ of cofinality 
$\om_n$, `` $\Theta_n$ has an $\alpha$-model " is independent of
$ZFC + GCH $.
\end{The}

\section{The local theories of $\omega_n$, $n\geq 1$}

\noi We have already mentioned in the introduction that it would be still possible 
that there are only finitely many possible local theories of $\om_2$ and that each of them 
is decidable, but that it is more plausible that the situation is much more complicated. 

On the other hand the above method cannot be applied to study the local theory of $\om_1$. 
We are going to prove in this section that the local theory of $\om_1$ is recursive in 
the local theory of $\om_2$, and more generally that, for all integers $n$, $p$, $1\leq n<p$, 
the local theory of $\om_n$ is recursive in 
the local theory of $\om_p$.  

\begin{Lem}\label{phin}
For each integer $n\geq 0$, there exists a local sentence $\vp_n$ such that 
$Sp_\infty(\vp_n)=[\om, \om_n]$. 
\end{Lem}

\proo  We have already proved this result in the cases $n=0$ and $n=1$ 
by proving Lemmas \ref{phi0} and \ref{phi1}.  We can now construct by induction on the integer 
$n$ a local sentence $\vp_n$ such that  $Sp_\infty(\vp_n)=[\om, \om_n]$.  The local sentence 
$\vp_n$ is constructed from the local sentence $\vp_{n-1}$ in a similar manner as   
in the construction of the local sentence $\vp_1$
from the local sentence $\vp_0$ (see the proof of  Lemma \ref{phi1}). 
Details are here left to the reader. 
\ep

\begin{Lem}\label{recS}
For every integer $n\geq 1$, there exists a recursive function 
$\mathcal{S}_n$ defined on the set of first order sentences (whose signatures contain 
the binary symbol $<$) 
such that,  for a first order sentence $\vp$,  
$[\vp \mbox{ is local }]$ if and only if 
$[\mathcal{S}_n(\vp) \mbox{ is local }]$ and 
$[\vp \mbox{ has an } \om_n$-model ] if and only if $[\mathcal{S}_n(\vp) 
\mbox{ has an } \om_{n+1}$-model ].  

\end{Lem}

\proo  Let $n$ be an integer $\geq 1$ and $\vp$ be a first order sentence 
in a signature S($\vp$). 
We are going to explain informally  the construction of the sentence  $\mathcal{S}_n(\vp)$ 
from the sentence $\vp$ using similar methods as in the preceding section. 

We can assume that S($\vp$)$\cap$ S($\vp_n$)=$\{ < \}$. 
The signature of $\mathcal{S}_n(\vp)$ is equal to 
S($\vp_n$) $\cup$  S($\vp$) $\cup$  $\{ s, t, R_1, R_2, R_3\}$, 
where $s$ is a new unary function symbol, $t$ is a new binary function symbol, and 
$R_1$, $R_2$, $R_3$ are three unary predicate symbols  not in 
S($\vp_n$) $\cup$  S($\vp$). 

The sentence $\mathcal{S}_n(\vp)$ expresses that a model $M$ is linearly ordered by the binary 
relation $<^M$, and that $R_1^M$, $R_2^M$, and $R_3^M$ are three successive segments of $M$. 
Then $\mathcal{S}_n(\vp)$ expresses that the restriction of $R_1^M$ 
to the signature of $\vp_n$ is a model of $\vp_n$ and the restriction 
of $R_2^M$ to the signature of $\vp$ is a model of $\vp$ (a k-ary function of S($\vp_n$) 
is trivially defined out of $R_1$ by $f(x_1, \ldots , x_k)=x_1$ and similarly functions of 
S($\vp$) are trivially defined out of $R_2$).  
\nl The function $s$ is a strictly non decreasing function from $R_2^M$ into $R_1^M$ and is 
trivially defined by $s(x)=x$ elsewhere. 
\nl The function $t$ is used to realize, for every element $a\in R_3^M$, an injection 
from $\{x\in R_3 \mid x<a\}$ into $R_2$ (as in the proof of Lemma \ref{phi1}) and 
it is trivially defined where it is not useful for that purpose. 

The function $\mathcal{S}_n$ is clearly recursive and it 
is easy to see that $\vp$ is local iff $\mathcal{S}_n(\vp)$ is local. 
\nl In that case it holds that $n_{ \mathcal{S}_n(\vp) }=  n_\vp + n_{\vp_n} + 2$. 
Indeed to take the closure of a set $X$ in a model $M$ of $\mathcal{S}_n(\vp)$ one takes 
the closure under the function $t$, then under the functions of S($\vp$) in $n_\vp$ steps, 
then under the function $s$, then under the functions of S($\vp_n$) in $n_{\vp_n}$ steps.  

Assume now that $\mathcal{S}_n(\vp)$ has an $\om_{n+1}$-model $M$. In this model 
$(R_1^M, <^M)$ and $(R_2^M, <^M)$ 
 have order types smaller than or equal to $\om_n$ because 
$Sp_\infty(\vp_n)=[\om, \om_n]$ and there is a 
strictly non decreasing function $s^M$ from $R_2^M$ into $R_1^M$. 
Thus $(R_3^M, <^M)$ must have order type $\om_{n+1}$. 
Every strict initial segment of $R_3^M$ is injected in $R_2^M$ so $R_2^M$ has cardinality 
$\aleph_n$ and its  order type is exactly $\om_n$. 
This implies that the restriction of the model   $M$ to $R_2^M$ and 
to the signature of $\vp$ is an $\om_{n}$-model 
of $\vp$. 
\nl Conversely it is easy to see that by construction  if there is an $\om_{n}$-model 
of $\vp$ then there is an $\om_{n+1}$-model of $\mathcal{S}_n(\vp)$. 
\ep 

We can now state the following result. Recall that the local theory of an ordinal 
$\alpha$ is the set of local sentences having a model of order type $\alpha$; it will be 
denoted by $LT(\alpha)$.  

\begin{The}\label{theorec}
For all integers $n, p \geq 1$, if $n < p$ then the local theory of $\om_n$ 
is recursive in the local theory of $\om_p$. 

\end{The}

\proo It follows directly from Lemma \ref{recS} that for each integer $n\geq 1$ the local theory 
of $\om_n$ is recursive in the local theory of $\om_{n+1}$ because 
$$LT(\om_n) = \mathcal{S}_n^{-1} ( LT(\om_{n+1} ) )$$
\noi where $\mathcal{S}_n$ is a recursive function. We can now infer, by 
induction on the integer $p > n$, that if $n < p$ then the local theory of $\om_n$ 
is recursive in the local theory of $\om_p$. 
\ep 

\begin{Rem}
We have called here local theory of $\alpha$ the set of {\bf all} local sentences 
having a model of order type $\alpha$. We could have restricted this set to local sentences 
in the recursive set {\bf L} given by Proposition \ref{rec}. 
\nl We can get a similar result in that case, defining firstly the recursive function 
$\mathcal{S}_n$ only on this set {\bf L} with values in {\bf L}. This is possible 
because we have seen that for a local sentence $\vp$ it holds that 
$n_{ \mathcal{S}_n(\vp) }=  n_\vp + n_{\vp_n} + 2$. Thus 
we can compute 
$n_{ \mathcal{S}_n(\vp) }$ from  $n_\vp$. 
\end{Rem}

Theorem \ref{theorec} states that  if $n < p$ then the local theory of $\om_n$ 
is less ``complicated" than  the local theory of $\om_p$ because there is a recursive reduction 
of the first one to the second one. 
\nl We are going to prove the following similar result. 

\begin{The}\label{recalpha}
For all integers $n \geq 1$, if $\alpha$ is an ordinal of cofinality $\om_n$ then 
the local theory of $\om_n$ 
is recursive in the local theory of $\alpha$. 
\end{The}

We shall proceed by successive lemmas.

\begin{Lem}\label{psi'n} 
Let $\psi$ be a local sentence and  $n$ be an integer $\geq 1$, then there exists 
 another local sentence $\psi_n'$ such that $Sp_\infty(\psi_n')\subseteq [\om, \om_n]$ and the 
following equivalence holds: 
[$\psi$ has a model of order type 
$\om_n$] iff [$\psi_n'$ has a model of order type $\om_n$].  
\end{Lem}

\proo  
Let $\psi$ be a local sentence (with $<\in$ S($\psi$)) and  $n$ be an integer $n \geq 1$.  
We now explain informally the construction of a local sentence $\psi_n'$
such that $Sp_\infty(\psi_n')\subseteq [\om, \om_n]$ and 
[$\psi$ has a model of order type 
$\om_n$] iff [$\psi_n'$ has a model of order type $\om_n$].  

The signature of $\psi_n'$ 
is 
S($\psi_n'$)=S($\vp_{n-1}$) $\cup$ S($\psi$) $\cup \{R, t\}$ 
where $R$ is a new unary 
predicate symbol and $t$ is a new binary function symbol not in S($\vp_{n-1}$) $\cup$ S($\psi$). 
The sentence $\psi_n'$ expresses that 
in a model $M$, $R^M$ is an initial segment of the model which is closed under functions of 
S($\vp_{n-1}$); and the restriction of $M$ to $R^M$ and to the signature S($\vp_{n-1}$) 
is a model of 
$\vp_{n-1}$. In the same way the  restriction of $M$ to $\neg R^M$ and to the signature 
 S($\psi$) is a model of $\psi$. The function $t$ 
is used to realize, for every element $a\in \neg R$, an injection 
from $\{x\in \neg R \mid x<a\}$ into $R$ (as in the proof of Lemma \ref{phi1}) and 
it is trivially defined where it is not useful for that purpose. 
\nl We know that the sentence $\vp_{n-1}$ given by Lemma \ref{phin} has infinite spectrum 
$Sp_\infty(\vp_{n-1})=[\om, \om_{n-1}]$ so in a well ordered model $M$ of $\psi_n'$ the initial 
segment $R^M$ will have order type $\leq \om_{n-1}$. Moreover every strict initial segment 
of $\neg R^M$ will be of cardinal $\leq \aleph_{n-1}$ because it is injected into $R^M$, so 
$\neg R^M$ will be of order type $\leq \om_{n}$ thus $M$ will 
be also of order type $\leq \om_{n}$. We have then proved that 
$Sp_\infty(\psi_n')\subseteq [\om, \om_n]$. 
\nl It is now easy to see that if $\psi_n'$ has a model $M$ of order type $\om_n$, then 
the restriction of $M$ to $\neg R^M$ and to the signature 
 S($\psi$) is a model of $\psi$ whose order type is $\om_n$; conversely if $\psi$ has an 
 $\om_n$-model then there is  an $\om_n$-model of $\psi_n'$. 

The sentence $\psi_n'$ is equivalent to a universal sentence and closure 
in its models takes at most 
$n_{\psi_n'}=n_\psi + 1 + n_{\vp_{n-1}}$. One takes closure under functions of 
S($\psi$) in $n_\psi$ steps,  then closure under the function $t$ in one step, then closure 
under functions of S($\vp_{n-1}$) in $n_{\vp_{n-1}}$ steps. 
Thus $\psi_n'$ is a local sentence.  \ep

\begin{Lem}\label{recT}
For every integer $n\geq 1$, there exists a recursive function 
$\mathcal{T}_n$,  defined on the set of first order sentences $\psi$ with $<\in$ S($\psi$), 
such that,  for every first order sentence $\psi$,  
$[\psi \mbox{ is local }]$ if and only if 
$[\mathcal{T}_n(\psi) \mbox{ is local }]$ and, for every local  sentence $\psi$,  
$[\psi \mbox{ has an } \om_n$-model $]$ if and only if $[\mathcal{T}_n(\psi) 
\mbox{ has an } \alpha$-model $]$ where $\alpha$ is any ordinal of cofinality $\om_n$.  

\end{Lem}

\proo  Let $n$ be an integer $\geq 1$ and $\psi$ be a first order sentence 
in a signature S($\psi$). We define $\mathcal{T}_n(\psi)=\theta^{\star \psi_n'}$ where 
$\theta$ is the local sentence whose spectrum is the class of successor ordinals, and 
$\psi_n'$ is the first order sentence constructed as above from the  sentence $\psi$. 
\nl Notice that in preceding lemma the sentence $\psi_n'$ is constructed from a {\it local 
sentence} $\psi$ but we can easily extend the construction to all {\it first order 
sentences} $\psi$. 
Then it holds that $\psi$ is local iff $\psi_n'$ is local. 
\nl The sentence $\theta^{\star \psi_n'}$ can also be defined even if 
$\psi_n'$ is not local, with slight modifications, in such a way that models of 
$\theta^{\star \psi_n'}$ are still essentially direct sums of models of $\theta$, these models 
being ordered by the order type of a model of $\psi_n'$. 
Moreover it holds also that [ $\theta^{\star \psi_n'}$ is local ] 
iff [ $\psi_n'$ is local ]. 
 Thus [ $\psi$ is local ] iff [ $\mathcal{T}_n(\psi)$=$\theta^{\star \psi_n'}$ is local ].
\nl Consider now a local  sentence $\psi$ and an ordinal 
$\alpha$ having  cofinality $\om_n$.  Then by Lemma \ref{psi'n} the sentence 
$\psi$ has an $\om_n$-model iff $\psi_n'$ has an  $\om_n$-model and 
$Sp_\infty(\psi_n')\subseteq [\om, \om_n]$. 
This implies that $[\psi \mbox{ has an } \om_n$-model $]$ iff $[ \theta^{\star \psi_n'} 
\mbox{ has an } \alpha$-model $]$ because $\alpha$ has cofinality $\om_n$.
\ep 

We can now end the proof of Theorem \ref{recalpha}. 
It follows from Lemma \ref{recT} that for each integer $n\geq 1$ the local theory 
of $\om_n$ is recursive in the local theory of $\alpha$, where  
$\alpha$ is an ordinal having  cofinality $\om_n$.  
Indeed 
$$LT(\om_n) = \mathcal{T}_n^{-1} ( LT(\alpha) )$$
\noi where $\mathcal{T}_n$ is a recursive function. 
\ep

\hs {\bf Acknowledgements.}
Thanks  to the anonymous referee for useful comments
on a preliminary version of this paper.

\end{document}